\documentclass[prd,showpacs,preprint,preprintnumbers,nobibnotes,superscriptaddress,floatfix]{revtex4}
\usepackage{amsmath,amssymb,amsbsy,color}
\usepackage{epsfig}
\usepackage{psfrag}
\usepackage[normalem]{ulem}

\usepackage{hyperref}

\begin{document}
\preprint{BNL-100706-2013-JA, RBRC-1016}

\title{Proton decay matrix elements on the lattice}

\author{Y. Aoki}
\email{yaoki@kmi.nagoya-u.ac.jp}
\affiliation{
  Kobayashi-Maskawa Institute for the Origin of Particles and the Universe (KMI), 
  Nagoya University, Nagoya 464-8602, Japan
}
\affiliation{
  RIKEN-BNL Research Center, Brookhaven National Laboratory, Upton, NY 11973, USA
}

\author{E. Shintani}
\email{shintani@riken.jp}
\affiliation{
  RIKEN-BNL Research Center, Brookhaven National Laboratory, Upton, NY 11973, USA
}

\author{A. Soni}
\email{adlersoni@gmail.com}
\affiliation{
  High Energy Theory Group, Brookhaven National Laboratory, Upton, NY 11973, USA
}

\collaboration{RBC and UKQCD collaborations}

\begin{abstract}

Hadronic matrix elements of proton decay are essential ingredients
to bridge the grand unification theory to low energy observables 
like proton lifetime.
In this paper we non-perturbatively calculate the 
matrix elements, relevant for the process of a nucleon decaying into 
a pseudoscalar meson and an anti-lepton through generic baryon number violating
four-fermi operators. 
Lattice QCD with 2+1 flavor dynamical domain-wall fermions
with the {\it direct} method, which is direct measurement of matrix element 
from three-point function without using chiral perturbation theory, 
are used for this study to have good control over the error due to 
lattice discretization effects, operator renormalization, and chiral extrapolation.
The relevant form factors for possible transition process from
an initial proton or neutron to a final pion or kaon induced by
all types of three quark operators are obtained through
three-point functions of (nucleon)-(three-quark operator)-(meson)
with physical kinematics.
In this study all the relevant systematic uncertainties 
of the form factors are taken into account for the first time, 
and 
the total error is found to be the range 30\%--40\%
for $\pi$ and 20\%--40\% for $K$ final states. 
\end{abstract}
\maketitle

\section{Introduction}

Proton decay is a smoking gun evidence of physics beyond the standard
model and is a natural outcome of Grand Unified Theories (GUTs) 
\cite{Pati:1973rp,Georgi:1974sy}.
The process occurs through baryon number changing interactions
mediated by the heavy new particles. 
Dominant modes are of X and Y gauge boson exchange for 
GUTs and of color-triplet Higgs multiplet for
supersymmetric (SUSY) GUTs \cite{Hisano:1992jj,Murayama:2001ur}.
Recent SuperKamiokande experiments report
the bound on proton partial lifetime, 
for instance, $\tau >8.2\times 10^{33}$ year for the 
$p\rightarrow e^+\pi^0$ channel 
\cite{Nishino:2009aa,Nishino:2012ipa}, which is typical for gauge boson
exchange, or $\tau>2.3\times 10^{33}$ for 
$p\rightarrow K^+\bar{\nu}$ \cite{Kobayashi:2005pe} and
$\tau>1.6\times 10^{33}$ for $p\rightarrow K^0\bar\mu^+$ \cite{Regis:2012sn},
both of which are favored for some SUSY GUTs.
There have been many arguments of a constraint on proton lifetime
from various types of GUT models so far 
%% \cite{Hisano:1992jj,Murayama:2001ur,Bajc:2002bv,Dorsner:2004xa,FileviezPerez:2004hn}
(see a comprehensive review \cite{Nath:2006ut} and reference therein).
In order to constrain the parameter space in GUT models 
with a reliable bound, 
a removal of all the theoretical uncertainties is highly desirable. 
One of the important elements, which can be made less uncertain
from the current knowledge, is the hadronic contribution to proton decay
matrix elements. 
Lattice QCD calculation can lead to reducing the uncertainties 
in the hadronic matrix element of a nucleon decaying 
into a pseudoscalar meson, and thus it can provide 
relevant information 
for the proton lifetime bound and help 
experimental plans for the future \cite{Abe:2011ts}.

The estimate of proton decay matrix elements in lattice QCD 
has been significantly improved by removing  
systematic errors, one by one,  since the first attempts in 1980s
\cite{Hara:1986hk,Bowler:1988us,Gavela:1989cp}.
A decade ago 
JLQCD collaboration \cite{Aoki:1999tw} 
performed an extensive calculation of proton decay matrix elements
using Wilson fermion action and operator renormalization estimated 
by one-loop lattice perturbation in the quenched approximation
with 
both the ``{\it direct}'' method, 
which is a direct measurement of matrix element from three-point functions,
and the ``{\it indirect}'' method,  
which is an effective estimate through low-energy constants 
in tree-level chiral perturbation theory, calculated
with two-point functions.
Few years later JLQCD and CP-PACS joint collaboration carried 
out a continuum extrapolation of the low energy 
constants for the {\it indirect} method \cite{Tsutsui:2004qc} to  
control the uncertainty of large discretization error.
Using the {\it direct} method,
RBC collaboration \cite{Aoki:2006ib}
performed the analysis 
with quenched domain-wall fermions (DWFs) and
non-perturbative renormalization, 
where thanks to almost exact chiral symmetry of the DWFs the discretization
error of $\mathcal O(a)$ is essentially removed and 
the error of renormalization factor associated with the use of
lattice perturbation theory was also eliminated.
The RBC collaboration also performed the DWF calculation using the
{\it indirect} method with quenched approximation
as well as with unquenching $u$ and $d$ quarks \cite{Aoki:2006ib}, and
later the RBC and UKQCD collaborations extended the DWF calculation of the 
{\it indirect} method using three dynamical quarks ($u$, $d$ and $s$)
\cite{Aoki:2008ku}.
In this way, one of the uncontrolled systematic uncertainty
coming from quenched approximation was removed.  

A striking, but perhaps not surprising outcome of the comparison
of the results from {\it direct} and {\it indirect} calculations,
though performed only with quenched approximation so far,
is that the {\it indirect} method could overestimate the matrix
elements by a factor of about two \cite{Aoki:2006ib}.
To fully control the systematic uncertainties, therefore, one needs to perform
the {\it direct} calculation with the $N_f=2+1$ dynamical simulations and
a non-perturbative operator renormalization.

In this paper we 
provide the non-perturbative estimate of proton decay matrix elements 
using the {\it direct} method with the dynamical, $N_f=2+1$
(degenerate $u$, $d$ and physical $s$ quarks) flavor lattice QCD with DWFs.
The DWF ensemble for $N_f=2+1$ at the lattice cutoff
$a^{-1}\sim 1.7$ GeV with 300--700 MeV pion masses \cite{Aoki:2010dy} 
in RBC/UKQCD collaboration are used for this purpose, and 
thus this enables us to evaluate hadronic matrix elements 
including almost all systematic errors on the lattice. 

This paper is organized as follows. 
In section \ref{sec:protonMA} we explain 
the definition and property of the 
matrix elements as well as their relation to the proton partial
decay width. 
The method to extract the matrix elements from three-point function on 
the lattice is expressed in section \ref{sec:latset}, and 
in section \ref{sec:result} we present our setup and the detailed analysis to 
obtain the matrix elements and evaluate their systematic uncertainties.
Section \ref{sec:summary} is devoted to summary and outlook.

\section{Proton decay matrix element}\label{sec:protonMA}

\subsection{Effective Lagrangian and matrix element}

Baryon number violating operators appearing in the leading low-energy 
effective Hamiltonian are constructed by possible combination of dimension-six 
(three quarks and one lepton) 
operators to be SU(3) color singlets and SU$_L$(2)$\times$ U$_Y$(1) invariant.
Following the notation of \cite{Weinberg:1979sa,Wilczek:1979hc,Abbott:1980zj}, 
four-fermi operators are expressed as
\begin{eqnarray}
 O^{(1)}_{abcd} 
   &=& (D_a^i,U_b^j)_R(q_c^{k\alpha},l_d^\beta)_L\varepsilon^{ijk}\varepsilon^{\alpha\beta}
   \label{eq:six-dimop1},\\
 O^{(2)}_{abcd} 
   &=& (q_a^{i\alpha},q_b^{j\beta})_L(U_c^{k},l_d)_R\varepsilon^{ijk}\varepsilon^{\alpha\beta}
   \label{eq:six-dimop2},\\
 \tilde{O}^{(4)}_{abcd}
   &=& (q_a^{i\alpha},q_b^{j\beta})_L(q_c^{k\gamma},l_d^\delta)_L\varepsilon^{ijk}
     \varepsilon^{\alpha\delta}\varepsilon^{\beta\gamma}
   \label{eq:six-dimop3},\\
 O^{(5)}_{abcd}
   &=& (U_a^{i},D_b^{j})_R(U_c^{k},l_d)_R\varepsilon^{ijk} \label{eq:six-dimop4},
\end{eqnarray}
with generic lepton field $l$, and quark field of left-handed part $q$ and 
right-handed part $U$ and $D$ as up and down type. 
The indices $a,b,c,d$ denote the generation number of fermion, $i,j,k$ denote
color SU(3) indices, and $\alpha,\beta,\gamma,\delta$ are SU(2) indices. 
The inner product is defined as $(x,y)_{R/L}=x^T CP_{R/L}y$ 
which has charge conjugation matrix $C$ and chiral projection $P_{R/L}$. 
The baryon number violation (but preserving $B-L$ number) 
in GUT models is generally expressed as 
low-energy effective Hamiltonian with the above six-dimension operators.
Leading term of effective Hamiltonian at low energies is represented as
\begin{eqnarray}
  \mathcal L^{B\hspace{-2mm}/} = \sum_I C^I\big[(qq)(ql)\big]^I + \cdots
  &=& -\sum_I C^I [\bar l^c\mathcal O_{qqq}\big]^I + \cdots,
  \label{eq:Heff}
\end{eqnarray}
where $C^I=C^I(\mu)$ is the Wilson coefficient with renormalization 
scale $\mu$ %and GUT scale $M_X$ 
of the operator $[(qq)(ql)]^I$ with $q$ being a light quark flavor $u$, $d$, or $s$.
The operator is one of those appearing
in Eq.(\ref{eq:six-dimop1})--(\ref{eq:six-dimop4}), and renormalized
also at $\mu$. 
The details of the  (SUSY) GUT is all captured in the coefficients $C^I(\mu)$.
Ellipsis means the higher order operators which are suppressed by
inverse power of heavy mass scale. 
The index $I$ distinguishes the type of operator with respect to
the quark-lepton flavor and chirality.
The three-quark operator reads
\begin{equation}
  \mathcal O^{\Gamma\Gamma'}_{qqq} = (qq)_\Gamma q_{\Gamma'} 
  = \varepsilon^{ijk}(q^{i\,T} CP_{\Gamma}q^j) P_{\Gamma'}q^k,
  \label{eq:OpNtoPS}
\end{equation}
% with light quark flavor $q=(u,d,s)$, 
where the color singlet contraction is taken.
Dirac spinor indices are omitted in the above equation.
In the following we may use simple notations for the three-quark operators
as $\mathcal O^{\Gamma\Gamma'}$.
$\Gamma$ and $\Gamma'$ denote the chirality, either $R$ or $L$ and 
the bracket means the contractions among Dirac spinors.

We calculate the transition matrix elements of the dimension-six 
operators with an initial nucleon (proton or neutron, $N=p,n$) state
and a final state containing a pseudoscalar meson ($P=(\pi,K,\eta)$)
and an anti-lepton ($\bar l$)
\begin{equation}
  \langle P(\vec p), l(\vec q,s)|[\bar l^c\mathcal O^{\Gamma\Gamma'}]|N(\vec k)\rangle
  = \bar v^c_{l}(q,s)\langle P(\vec p)|\mathcal O^{\Gamma\Gamma'}|N(\vec k,s)\rangle,
  \label{eq:NtoLPS}
\end{equation}
including three-dimensional momenta, $\vec p$ for final pseudoscalar, 
$\vec k$ for initial nucleon and $\vec q = \vec p - \vec k$ for final lepton
which is determined from momentum conservation.
Neglecting the electroweak interaction of the lepton,
the amplitude $\langle l(\vec q,s)|\bar l^c|0\rangle = \bar v_l^c(\vec q,s)$ 
of the lepton part can be captured in the wave function 
of on-shell lepton state at momentum $\vec q$ for spin $s$ component.
The matrix element 
$\langle P(\vec p)|\mathcal O^{\Gamma\Gamma'}|N(\vec k,s)\rangle$ 
is parametrized by the relevant form factor $W_0(q^2)$ and irrelevant
one $W_1(q^2)$ as 
\begin{equation}
  \langle P(\vec p)|\mathcal O^{\Gamma\Gamma'}|N(\vec k,s)\rangle
  = P_{\Gamma'}\Big[ W^{\Gamma\Gamma'}_0(q^2) 
    - \frac{iq\hspace{-2mm}/}{m_N}W^{\Gamma\Gamma'}_1(q^2)\Big]u_N(k,s). \label{eq:NtoPS}
\end{equation}
$W_0$ and $W_1$ are defined for each matrix element with the three-quark operator
renormalized in $\overline{MS}$ NDR at scale $\mu$, and are functions of 
square of four momentum transfer $q=k-p$. 
Using on-shell condition, the total matrix element as shown in Eq.(\ref{eq:NtoLPS}) 
is given by
\begin{eqnarray}
  \bar v^c_{l}(q,s)
    \langle P(\vec p)|\mathcal O^{\Gamma\Gamma'}|N(\vec k,s)\rangle
&=& \bar v_l^c(q,s) P_{\Gamma'}
    \Big[ W^{\Gamma\Gamma'}_0(q^2) -\frac{iq\hspace{-2mm}/}{m_N}W^{\Gamma\Gamma'}_1(q^2)\Big]
    u_N(k,s)\nonumber\\
&=& \bar v^c_l(\vec q,s) P_{\Gamma'}u_N(\vec k,s) W^{\Gamma\Gamma'}_0(0) + O(m_l/m_N),
\label{eq:NtoPSelem}
\end{eqnarray}
with $iq\hspace{-2mm}/v_l = m_l v_l$ and $W_1\simeq W_0$ \cite{Aoki:2006ib}. 
Since $-q^2 = m^2_l$ is much smaller than nucleon mass squared in the case 
of $l=e,\nu$, we set $q^2=0$ and ignore the second term in Eq.(\ref{eq:NtoPSelem}).
Taking only the relevant form factor will be a good approximation
even for $l=\mu$, as $m_\mu/m_N\sim 10$\% is smaller than the total error of
$W_0$ in this study.

Once the relevant form factor $W_0$ is obtained in lattice QCD,  
the partial decay width of the decay $N\rightarrow P + \bar l$ is given by 
\begin{equation}
  \Gamma(N\rightarrow P+\bar l) 
   = \frac{m_N}{32\pi}\Big[ 1 - \Big(\frac{m_{P}}{m_N}\Big)^2\Big]^2
  \Big| \sum_I C^I W_0^I(N\rightarrow P)\Big|^2
\end{equation}
with the perturbative estimate of Wilson coefficient 
$C^I$ in the GUT models \cite{Nath:2006ut}.
Note that renormalization scale dependence of $C^I$ and $W_0^I$ cancels out 
in their multiplication. 

The different chirality combinations of the matrix elements are related 
through the Parity transformation as
\begin{eqnarray}
  &&\langle P; \vec p|\mathcal O^{R L}|N; \vec k,s\rangle
  = \gamma_0\langle P; -\vec p|\mathcal O^{L R}|N; -\vec k,s\rangle,\\
  &&\langle P; \vec p|\mathcal O^{L L}|N; \vec k,s\rangle
  = \gamma_0\langle P; -\vec p|\mathcal O^{R R}|N; -\vec k,s\rangle,
\end{eqnarray}
which indicates that four chirality combinations $(\Gamma\Gamma')=(RL),(LL),(LR),(RR)$
are reduced to two different combinations, $(\Gamma\Gamma')=(RL),(LL)$.
In the following $\Gamma'$ is fixed in a left-handed chirality, and 
a short-hand notation $W_{0,1}^{\Gamma L}\equiv W_{0,1}^{\Gamma}$ is used.
Under exchange-symmetry between $u$ and $d$ 
there are the following relations between proton and neutron matrix elements:
\begin{eqnarray}
  \langle \pi^0| (ud)_\Gamma u_L | p\rangle &=& \langle \pi^0| (du)_\Gamma d_L | n\rangle,
  \label{eq:n_p_1}\\
  \langle \pi^+| (ud)_\Gamma d_L | p\rangle &=& -\langle \pi^-| (du)_\Gamma u_L | n\rangle,
  \label{eq:n_p_2}\\
  \langle K^0| (us)_\Gamma u_L | p\rangle &=& -\langle K^+| (ds)_\Gamma
   d_L | n\rangle, \label{eq:n_K0}\\
  \langle K^+| (us)_\Gamma d_L | p\rangle &=& -\langle K^0| (ds)_\Gamma u_L | n\rangle,\\
  \langle K^+| (ud)_\Gamma s_L | p\rangle &=& -\langle K^0| (du)_\Gamma s_L | n\rangle,\\
  \langle K^+| (ds)_\Gamma u_L | p\rangle &=& -\langle K^0| (us)_\Gamma d_L | n\rangle,\\
  \langle \eta| (ud)_\Gamma u_L | p\rangle &=& -\langle \eta| (du)_\Gamma d_L | n\rangle.
  \label{eq:n_eta}
\end{eqnarray}
A negative sign comes from the interpolation operator of proton or
neutral pion by the exchange of $u$ and $d$.
Furthermore in the SU(2) isospin limit there is an additional relation between 
Eq.(\ref{eq:n_p_1}) and Eq.(\ref{eq:n_p_2}):
\begin{equation}
  \langle \pi^0| (ud)_\Gamma u_L | p\rangle = \sqrt 2\langle \pi^+| (ud)_\Gamma d_L | p\rangle.
  \label{eq:piplus}
\end{equation}
Therefore there are twelve principal matrix elements we calculate in this paper.

\section{Calculation scheme for the form factors}\label{sec:latset}

To obtain the matrix element we make use of the ratio of three-point function 
of (proton)-($O^{\Gamma L}$)-(meson) and two-point function of nucleon and meson. 
Such a ratio is represented as
\begin{eqnarray}
&& R_3(t,t_1,t_0;\vec p,\mathcal P) \nonumber\\
&& = \frac{ \sum_{\vec x,\vec x_1} e^{i\vec p(\vec x_1-\vec x)} {\rm tr}\big[ 
      \mathcal P\langle 0| J^{\rm gs}_{P}(\vec x_1,t_1) 
      \mathcal O^{\Gamma L}(\vec x,t) \bar J^{\rm gs}_p(\vec 0,t_0)|0\rangle \big] }
    { \sum_{\vec x,\vec x_1}e^{i\vec p(\vec x_1-\vec x)}\langle 0| J^{\rm gs}_{P}(\vec x_1,t_1) 
      J^{{\rm gs}\,\dag}_{P}(\vec x,t)|0\rangle\,
    \sum_{\vec x}{\rm tr}[P_4\langle 0|J^{\rm gs}_{p}(\vec x,t)
    \bar J^{\rm gs}_{p}(\vec 0,t_0)|0\rangle]}
    \sqrt{Z_{P}^{{\rm gs}}(\vec p)Z^{\rm gs}_p}L_\sigma^3,\nonumber\\
  \label{eq:ratio}
\end{eqnarray}
with interpolating field for pseudoscalar $J^{\rm gs}_{P}$ and proton $J^{\rm gs}_p$.
These interpolating operators are made of quark fields smeared using the gauge-invariant
Gaussian smearing \cite{Alexandrou:1992ti} with the parameters optimized for meson 
and proton separately. 
In the periodic lattice the injected spatial momentum is
$\vec p=2\pi \vec n/L_\sigma$, 
where $\vec n$ is integer vector $0\le n_i\le L_\sigma-1$, 
and $L_\sigma$ is the spatial extension of the lattice. 
``tr'' represents trace over spinor indices, and $\mathcal P$ is a spin projection matrix.
The three-point function in numerator is constructed by 
quark propagator with the sequential source method at pseudoscalar sink
location.

% In this calculation we use $\vec n=(1,0,0),(1,1,0)$ for meson momentum and 
% zero nucleon momentum in three-point function. 
$Z_{P,p}$ indicates the amplitude of overlap of the interpolating field 
to on-shell state,
\begin{eqnarray}
  \langle P(\vec p)| J_{P}^{\rm gs\,\dag}(0) | 0\rangle &=& \sqrt{Z^{\rm gs}_{P}(\vec p)},\\
  \langle 0 | J^{\rm gs}_{p}(0) | p(\vec 0,s) \rangle &=& \sqrt{Z^{\rm gs}_{p}}u_p(k,s),
\end{eqnarray}
with the proton Dirac spinor normalized as $\bar u_p(k,s)u_p(k,s') = 2m_N\delta_{ss'}$.
In this study we always take the proton to be at rest. 
%being the momentum of the nucleon $k=(0,0,0,im_N)$ and meson $p=(\vec{p},iE_{P})$.
Note that the operator of nucleon interpolating field is not uniquely determined, and 
we make use of the two possible proton operators formed as 
\begin{eqnarray}
 J_p = \varepsilon^{ijk}(u^{i\,T} C\gamma_5 d^j)u^k,\quad 
       \varepsilon^{ijk}(u^{i\,T}C\gamma_4\gamma_5 d^j)u^k.
\end{eqnarray} 
% Since there is equivalence between proton and neutron matrix element 
% as shown in Eq.(\ref{eq:n_p_1})--(\ref{eq:n_eta}), we focus on proton case.
Numerical comparison between the above two types of nucleon interpolating operator
will be shown in the next section.

In the simulation we take the sufficiently large separation between 
$t_0$ and $t_1$ in Eq.(\ref{eq:ratio}) so we have a range of $t$ 
where the three and two point functions 
in the ratio are dominated by the ground states.
Then the ratio leads to its asymptotic form, 
\begin{equation}
  \lim_{t_1-t,t-t_0\rightarrow\infty}R_3(t,t_1,t_0;\vec p,\mathcal P) 
  = R^{\rm asym}_3(\vec p,\mathcal P) 
  = {\rm tr}\Big[\mathcal PP_L\big(W_0^{\Gamma}(q^2) 
    - \frac{iq\hspace{-2mm}/}{m_N}\,W_1^{\Gamma}(q^2)\big)\Big],
\end{equation}
where $q^2$ is the squared momentum transfer from the initial proton
to the final pseudoscalar meson state $q^2=(k-p)^2$. 
We employ two different projection matrices $\mathcal P=P_4$ or $iP_4\gamma_j$
with $P_4=(1+\gamma_4)/2$ 
to subtract the contribution from the parity partner of the 
proton and to disentangle $W_0$ and $W_1$.
By solving the linear equations, 
% by which we subtract parity-partner contamination (excited state)
% to extract form factor from three point function by solving 
% linear equation of
\begin{eqnarray}
  &&R^{\rm asym}_3(p,P_4) 
  = W_0^{\Gamma}(q^2) - \frac{iq_4}{m_N}W_1^{\Gamma}(q^2),\label{eq:prj_I}\\
  &&R^{\rm asym}_3(p,iP_4\gamma_j) 
  = \frac{q_j}{m_N}W_1^{\Gamma}(q^2).\label{eq:prj_gmj}
\end{eqnarray}
the relevant form factor $W_0$ can be obtained.
% Hereafter we concentrate on the relevant form factor $W_0$.

\section{Numerical calculation of the proton decay form factors
 }\label{sec:result}

\subsection{Lattice setup}
We use the gauge configurations generated for $2+1$ flavor dynamical 
domain-wall fermions with Iwasaki gauge action by RBC and UKQCD 
collaborations \cite{Aoki:2010dy}. The lattice volume is $24^3\times 64$ 
and the size of the fifth dimension is $L_s=16$. 
The gauge coupling $\beta=2.13$ corresponds to $a^{-1}=1.73(3)$ GeV.
% We use with dynamical domain-wall fermions 
% of $2+1$ flavor at Iwasaki gauge action in lattice size $24^3\times 64$ 
% at $\beta=2.13$ which corresponds to $a^{-1}=1.73(3)$ GeV \cite{Aoki:2010dy}. 
This is the same ensemble as the previous {\it indirect} method study \cite{Aoki:2008ku}.
Boundary condition is periodic for the gauge field, and spatially periodic and 
temporally anti-periodic for the fermion fields.
We use four different unitary $u$, $d$ quark masses for chiral
extrapolation, and 
one unitary and one partially quenched strange-quark mass for the study of 
strange quark mass dependence for final $K^{0,+}$ kaon state.
For later convenience let us introduce the quark mass $\tilde{m}$
which includes the additive renormalization due to the
inexact chiral symmetry of the domain-wall fermions at a finite
extent of the fifth dimension. 
We define 
\begin{equation}
 \tilde{m} = m + m_{\rm res},
 \label{eq:addm}
\end{equation}
as the multiplicatively renormalizable mass with $m$ in the lattice action, 
where residual mass $m_{\rm res}$ for the lattice used in this study has been
calculated as $m_{\rm res}=0.003152(43)$ \cite{Aoki:2010dy}.
The form factors of the nucleon to pion matrix elements depend 
on $\tilde{m}_{ud}$ for the degenerate $u$ and $d$ quark mass
and the squared momentum transfer $q^2$. 
For the nucleon to kaon matrix elements, the strange quark mass $\tilde{m}_s$ 
enters as an additional parameter.

In the computation of the two-point and three-point function on the lattice, 
we employ a gauge-invariant Gaussian smearing with the optimized parameter
$(n_G,\sigma) = (40,5.0)$ for baryon source/sink and $(n_G,\sigma) = (16,3.0)$ 
for meson sink, where the APE-smeared gauge links with $(N,c)=(12,0.4)$
as defined in \cite{Lin:lat2012}.
The time slices for the nucleon source $t_0$ and meson sink $t_1$ are set
as $(t_0,t_1)$ = (5,37) or (27,59). The baryon number violating operator
at time $t$ moves between them ($t_0<t<t_1$).
We use first and second smallest but non-zero momentum
$p=(\pi/12,0,0)$, $(\pi/12,\pi/12,0)$ on the periodic lattice for the meson.
The statistics used for each ensemble is summarized in Table \ref{tab:latprm},
as well as with the used valence masses and the measured $q^2$.
Measurements are done with each 40 HMC trajectories for the ensembles
with $m_{ud}=0.005$ and 0.01, or 20 HMC trajectories for $m_{ud}=0.02$ and 0.03.
We alternate the source time slice $t_0$=5 and 27 from the one to the next
configuration for $m_{ud}=0.01$, 0.02 and 0.03, while we measure both
$t_0=$5 and 27 for all configurations at $m_{ud}=0.005$ (therefore the
number of measurements is doubled the number of configurations).
% Nucleon source-point $t_0$ and pseudoscalar sink-point $t_1$ for three-point function 
% are fixed at $t_0=5$ or 37 and $t_1=27$ or 59 with 
% operator at three-point function 
% constructed by sequential source method.  
% The meson sink moves its position $t$ 
% between $t_0$ and $t_1$ with 
% spacial momentum $\vec p=(\pi/12, 0, 0), (\pi/12,\pi/12,0)$.
%% The momentum of meson two-point function in the denominator of Eq.(\ref{eq:ratio}) is 
%% $\vec p=(\pm\pi/12, 0, 0), (\pm\pi/12,\pm\pi/12,0)$
%% and we take average over each signs. 
% The detailed number of gauge ensembles and source positions employed here 
% are shown in table \ref{tab:latprm}.
% For measurement at $m_{ud}^{\rm sea}=0.005$ ensemble we add the statistics in 
% different source and sink points in the same configurations separated by
% 40 HMC trajectories.
% At $m_{ud}^{\rm sea}=0.02, 0.03$ 
% the skipped HMC trajectory in the same source 
% and sink point is 40 and in every source-sink combinations we 
% separated 20 HMC trajectories
% in order to reduce the autocorrelation effect.
% At $m_{ud}^{\rm sea}=0.01$ the skipped HMC trajectory is 80 and  
% separated trajectory for every source and sink combinations is 40.
% Note that at the lightest quark mass ($m_{ud}^{\rm sea}=0.005$) 
% to increase statistics we further take average over
% two different source time slices, $t_0=$ 5 and 37, 
% for each gauge configurations therefore the number of measurements
% is double the number of configurations.

The multiplicative renormalization factors to convert the lattice 
three-quark operators in Eq.(\ref{eq:n_p_1})--(\ref{eq:n_eta}) 
into those in $\overline{MS}$ NDR scheme
has been calculated through the RI/MOM 
non-perturbative renormalization \cite{Aoki:2008ku} as
\begin{eqnarray}
 U(\mu=2{\rm GeV})_{LL} &=& 0.662(10)(53),\label{eq:uLL}\\
 U(\mu=2{\rm GeV})_{RL} &=& 0.665(8)(53)\label{eq:uRL}.
\end{eqnarray}
The first error is statistical one and the second is systematic one
(systematic error of 8\% is estimated in \cite{Aoki:2008ku}
as a truncation effect of the perturbative expansion). 

\begin{figure}[tb]
\begin{center}
  \includegraphics[width=100mm]{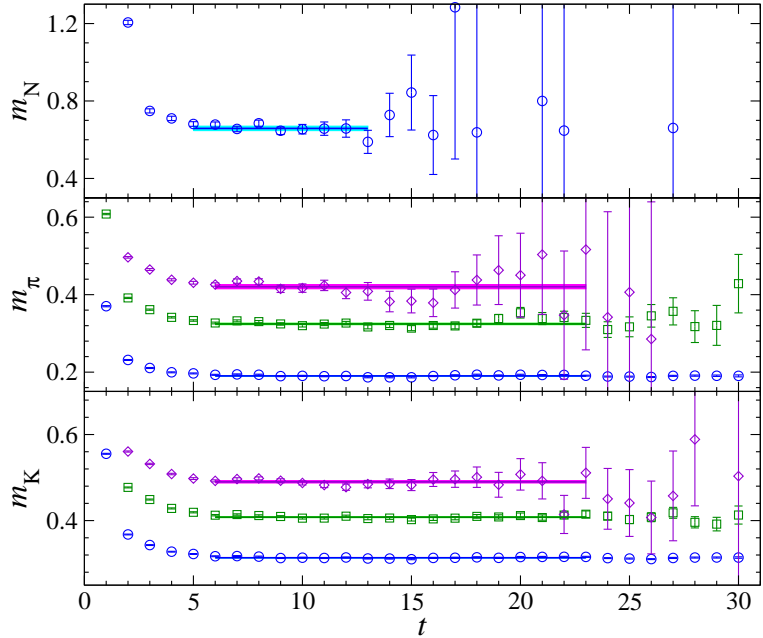}
  \vskip 3mm
  \caption{Effective mass plot of nucleon (top), pion (middle) and Kaon (bottom)
  at momentum square $n_p^2=0$ (circle), $n_p^2=1$ (square), $n_p^2=2$ (diamond)
  which correspond to $\vec p=(0, 0, 0), (\pi/12, 0, 0), (\pi/12,\pi/12,0)$ 
  respectively.
  For nucleon we use gauge-invariant Gaussian source/sink, and for 
  meson we use (Kuramashi-)wall source and gauge-invariant Gaussian sink.
  This is for the lightest quark mass $m_{ud}=0.005$ and $m_{s}=0.0343$.
  Solid line (colored band) indicate the central value (statistical error) 
  obtained by fitting. 
  }
  \label{fig:effm}
\end{center}
\end{figure}

In Figure \ref{fig:effm} we show the effective mass of nucleon, pion and 
kaon two-point function which enter in the denominator of Eq.(\ref{eq:ratio}).
The effective mass at time $t$ is constructed with data at $t$ and $t+1$, and 
we can observe the plateau region whose starting point is $t=5$ for the nucleon and 
$t=6$ for the pseudoscalar. 
Therefore, the denominator of Eq.(\ref{eq:ratio}) is dominated by 
the ground states for $t$ satisfying both $t-t_0\ge 5$ and $t_1-t\le 6$.
% It turns out that the asymptotic state of nucleon and meson state 
% of three point function in Eq.(\ref{eq:ratio}) 
% will be reached in the region $t-t_0\ge 5$ and $t_1-t\le 6$ for 
% temporal position of operator $t$.

\begin{table}
\caption{
 Lattice parameters, the estimate of the hadron masses and the squared
 momentum transfer from the initial state nucleon to the final state meson for
 each parameter set are shown. The lines with blank $m_s^{\rm val}$ entry
 show the kinematic parameters for the pion final state and nucleon mass,
 while those with $m_s^{\rm val}$ entry for the kaon final states. Two $-q^2$
 values in each line are for the two different momenta injected to the meson,
 $\vec p^2=(\pi/12)^2$, $2(\pi/12)^2$ respectively, where the $-q^2$ is shown in GeV unit
 using $a^{-1}=1.73(3)$ GeV \cite{Aoki:2010dy}.  Fitting range used for the mass estimate are
 $6\le t\le 23$ for pion and kaon or $5\le t\le 13$ for nucleon.
}\label{tab:latprm}
\begin{tabular}{ccccccccccc}
\hline\hline
    $(m_{ud}^{\rm sea},m_s^{\rm sea})$ 
  & $m_{ud}^{\rm val}$ 
  & $m_{s}^{\rm val}$ 
  & $m_{\pi}$ & $m_{K}$ & $m_N$ 
  & $-q^2$(GeV$^2$) &
  & \# configs.
  & \# meas.\\
\hline
 (0.005,0.04) & 0.005 &        & 0.1897(5) &   & 0.656(16) & $-$0.129 & 0.241 & 202
              & 404\\
              & 0.005 & 0.0343 &  & 0.3131(5) &     & 0.017 & 0.325 \\
              & 0.005 & 0.04   &  & 0.3322(5) &     & 0.039 & 0.337 \\
\hline
 (0.01,0.04)  & 0.01  &        & 0.2420(6) &     & 0.705(16) & $-$0.162 & 0.194 & 150
              & 150\\
              & 0.01  & 0.0343 &     & 0.3328(6) &     & $-$0.035 & 0.280 \\
              & 0.01  & 0.04   &     & 0.3510(6) &     & $-$0.011 & 0.295 \\
\hline
 (0.02,0.04)  & 0.02  &        & 0.3228(6) &     & 0.790(10) & $-$0.218 & 0.137 & 100
              & 100\\
              & 0.02 & 0.0343  &     & 0.3681(6) &     & $-$0.142 & 0.189\\
              & 0.02 & 0.04    &     & 0.3849(6) &     & $-$0.114 & 0.208 \\
\hline
 (0.03,0.04)  & 0.03 &         & 0.3880(7) &     & 0.912(11) & $-$0.391 & $-$0.020 & 90
              & 90\\
              & 0.03 & 0.0343  &     & 0.4003(6) &     & $-$0.364 & $-$0.000 \\
              & 0.03 & 0.04    &     & 0.4160(6) &     & $-$0.330 &  0.025\\
\hline\hline
\end{tabular}
\end{table}

\subsection{Measurement of the form factor and kinematics}

\begin{figure}[tb]
\begin{center}
  \includegraphics[width=120mm]{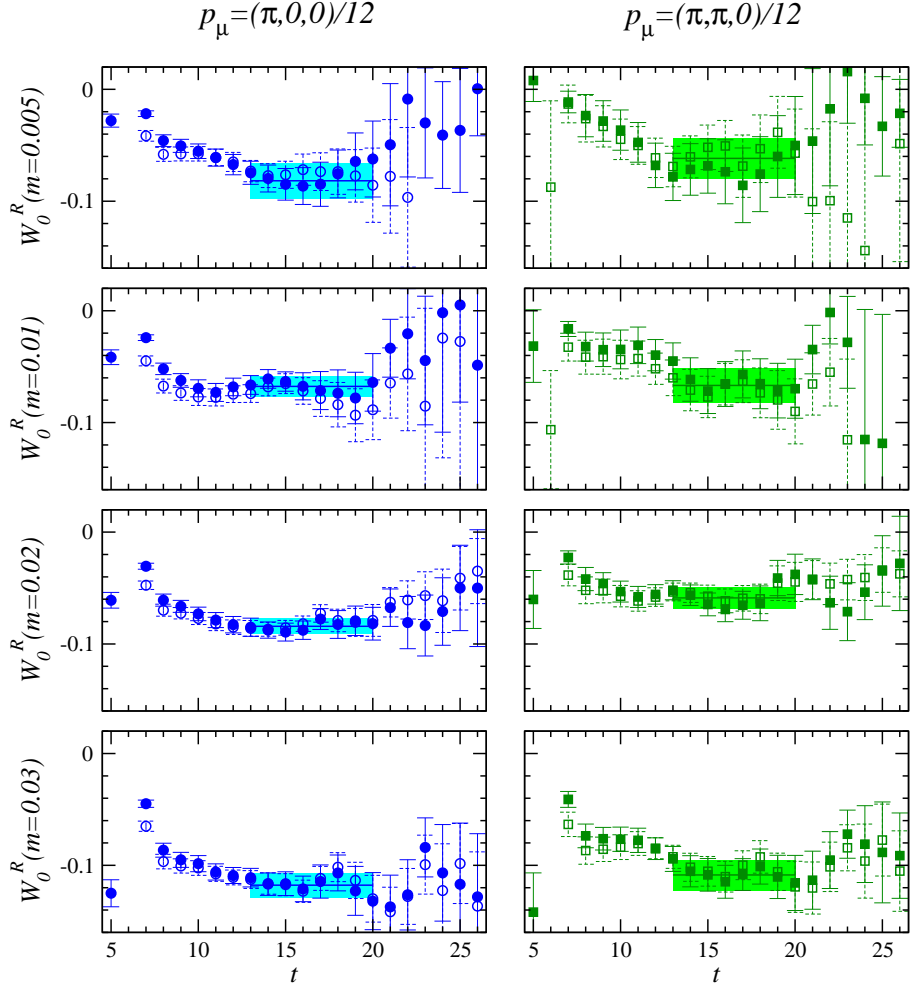}
  \vskip 3mm
  \caption{$W_0^R$ for $p\rightarrow \pi^0$ decay channel is plotted 
  as a function of operator time ($t$ in Eq.~(\ref{eq:ratio})).
  The proton source is located at $t=5$, and the $\pi^0$ sink is at $t=27$.
  Different symbols show the two different proton interpolating fields, 
  which correspond to $(u^T C\gamma_5d)u$ (open) and $(u^T C\gamma_4\gamma_5d)u$ (filled).
  The horizontal solid line indicates the central value of constant fit 
  to the both plateaus in the range $13\le t\le 20$ simultaneously.
  The shaded area indicates 1-sigma error band. 
  }
  \label{fig:W0_t_R}
\end{center}
\end{figure}

\begin{figure}[tb]
\begin{center}
  \includegraphics[width=120mm]{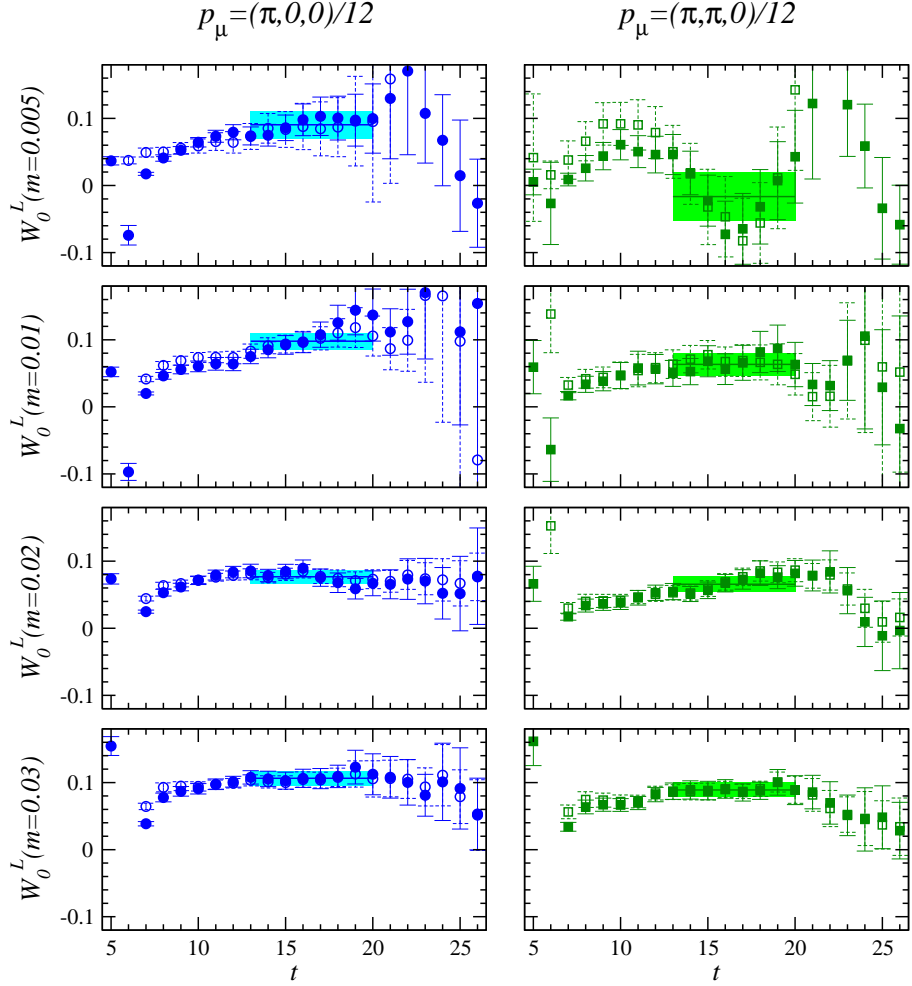}
  \vskip 3mm
  \caption{$W_0^L$ for $p\rightarrow \pi^0$ decay channel is plotted 
 as a function of operator time.
  Symbols are same as in Figure \ref{fig:W0_t_R}.}
  \label{fig:W0_t_L}
\end{center}
\end{figure}

Figures \ref{fig:W0_t_R} and  \ref{fig:W0_t_L} show 
the form factor $W_0$ of the $p\rightarrow \pi^0$ channel 
in Eqs.~(\ref{eq:prj_I}) and (\ref{eq:prj_gmj}) 
as a function of the time position $t$ of the three-quark operator.
The open and filled symbols correspond to results in two different 
nucleon interpolating operators,
$(q^T C\gamma_5 q)q$ and $(q^T C\gamma_4\gamma_5 q)q$ respectively.
To obtain the value of $W_0$, a simultaneous fit of these 
two effective $W_0$ is performed at the plateau
in the range $13\le t \le20$, where the two $W_0$ 
appear to be consistent and the contamination from
the excited states dies out. 
The same range is used for all the parameters and all the matrix elements.
Figures \ref{fig:W0_q2_R} and \ref{fig:W0_q2_L} show $W_0^{R/L}$ 
for each channel as a function of $q^2$.

The form factors in the physical kinematics are calculated from
the extrapolation or interpolation with momentum and quark masses. 
For the physical kinematics of
proton decay into meson and lepton final state, 
$-q^2$ is equivalent to lepton mass squared 
in the relevant form factor $W_0(q^2)$. 
In the lattice computation, however, the quark masses are other parameters that need
to be tuned toward the physical pion and kaon masses.
Therefore we have three parameters to tune: 
degenerate $u$, $d$ quark mass $\tilde m_{ud}$, strange quark mass $\tilde m_s$ 
and meson momentum $|\vec{p}|$.
In our simulation, the $\tilde{m}_{ud}\rightarrow \tilde{m}_{ud}^{\rm phys}$
limit is taken by an extrapolation, $\tilde{m}_{s}\rightarrow \tilde{m}_{s}^{\rm phys}$ 
limit is taken by an interpolation,
where physical quark mass 
in lattice units is realized by the limit,
\begin{eqnarray}
 \tilde{m}_{ud}^{\rm phys}&=&0.001385,\label{eq:mudphys}
 \\
 \tilde{m}_s^{\rm phys}&=&0.03785,\label{eq:msphys}
\end{eqnarray}
with the values to reproduce the experimental
hadron mass ratios, $m_\pi/m_\Omega$ and $m_K/m_\Omega$,
the pion and kaon mass over the mass of $\Omega^{-}$ \cite{Aoki:2010dy}.

We employ two different procedures for taking the above limit.
One is the global fit with a function that depends on both
quark mass and $q^2$, and thus 
$W_0$ at physical point is straightforwardly obtained.
The other is to sequentially take the two limits; first 
$q^2\to 0$ and then take the quark mass to the physical point.
In this procedure $W_0$ at physical point is obtained by the second limit.
In the next section we will show numerical results with these procedures.

\subsection{Extrapolation to physical kinematics with global fitting}\label{seq:global}

\begin{figure}[tb]
\begin{center}
  \vskip -30mm
  \includegraphics[width=170mm]{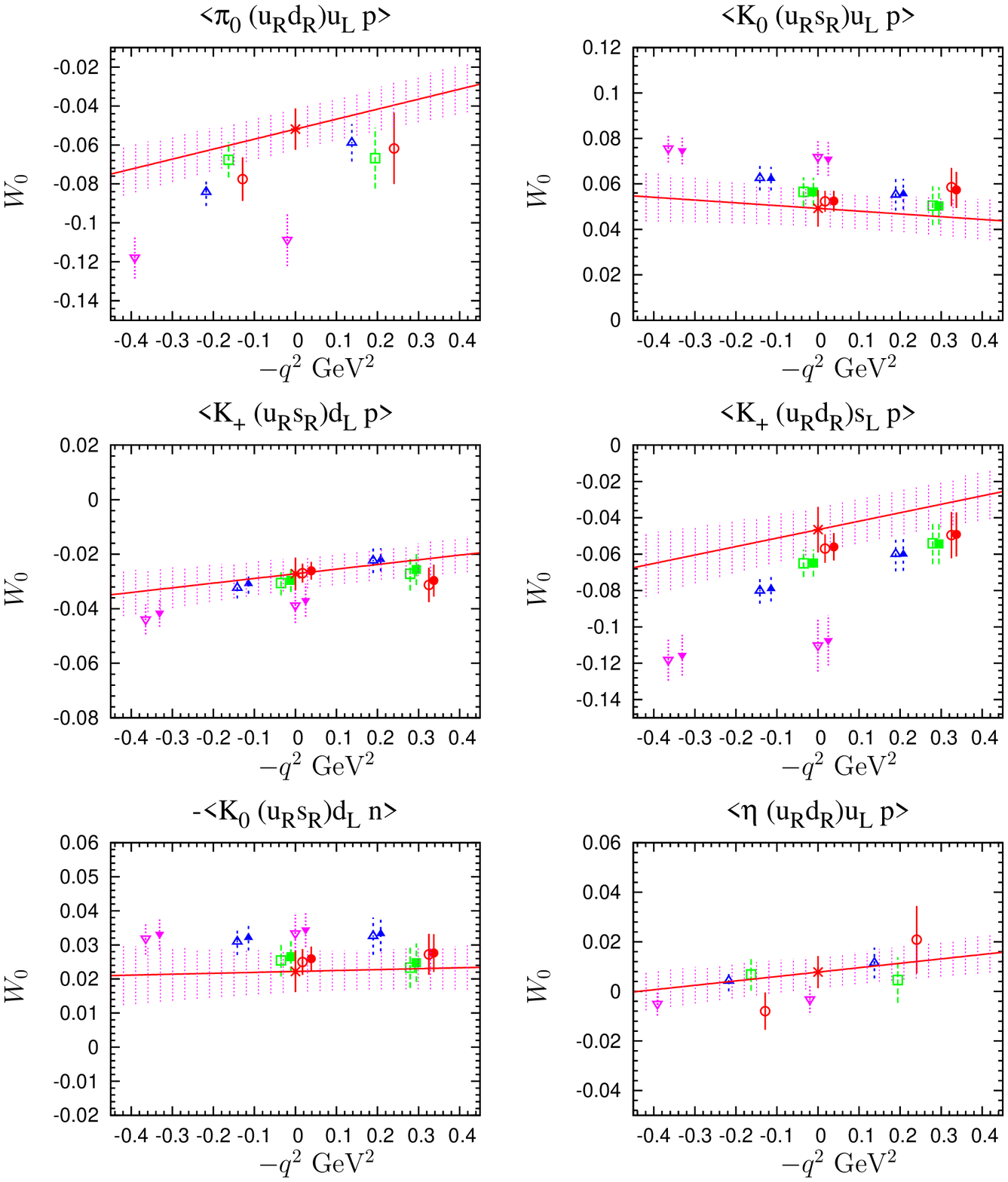}
  \vskip -20mm
  \caption{$q^2$ dependence of $W_0^R(q^2)$ at all quark masses in lattice units. 
  We plot the results at $m_{ud}=0.005$ (circle), $m_{ud}=0.01$ (square), 
  $m_{ud}=0.02$ (up-triangle) and $m_{ud}=0.03$ (down-triangle).
  In the figure for $K^{0,+}$, results at $m_{s}=0.0343$ represent open symbol
  and filled symbol at $m_{s}=0.04$.
  The solid lines (bands) show the global fit function (and its error)
  after taking the extrapolation into the physical quark mass
  using all of the points. 
  The star symbol is the result at the
  physical kinematics using the global fit.
  }
  \label{fig:W0_q2_R}
\end{center}
\end{figure}

\begin{figure}[tb]
\begin{center}
  \vskip -30mm
  \includegraphics[width=170mm]{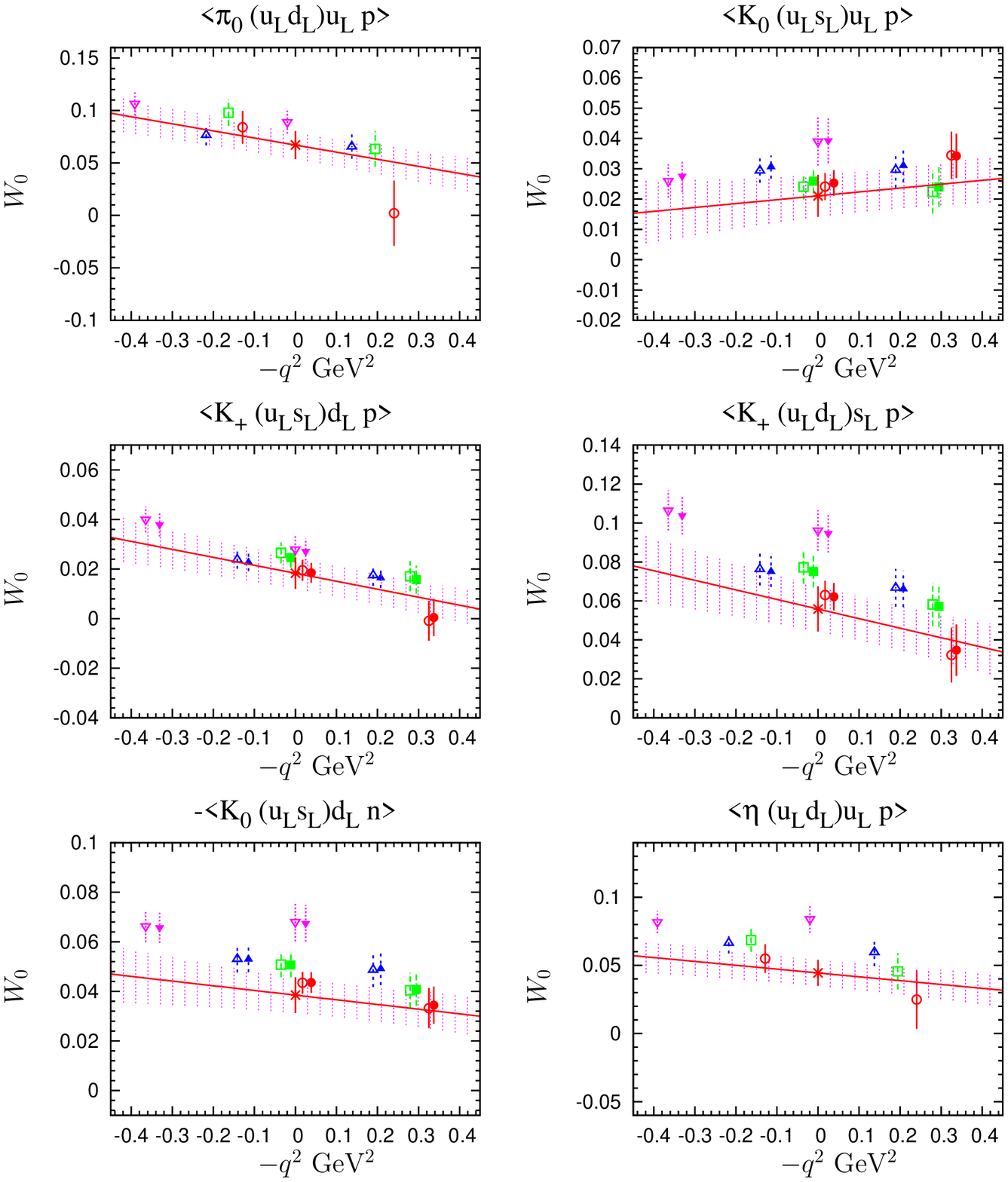}
  \vskip -20mm
  \caption{$q^2$ dependence of $W_0^L(q^2)$ at all quark masses. 
  Symbols are same as Figure \ref{fig:W0_q2_R}.
  }
  \label{fig:W0_q2_L}
\end{center}
\end{figure}

In the global fitting to obtain the form factor in the physical kinematics
we use the ansatz of linear function,
\begin{eqnarray}
  F_{W_0}^{\pi,\eta}(\tilde{m}_{ud},q^2) &=& A_0 + A_1 \tilde{m}_{ud} +
   A_2q^2,\label{eq:fit1_sim}\\ 
  F_{W_0}^K(\tilde{m}_{ud},\tilde{m}_s,q^2) &=& B_0 + B_1 \tilde m_{ud} + B_2
   \tilde{m}_s + B_3q^2,\label{eq:fit2_sim}
\end{eqnarray}
with free parameters $A_i$ and $B_i$.  $F_{W_0}^{\pi,\eta}$ is used for the
pion or $\eta$ final state, $F_{W_0}^K$ for the kaon final state.
This procedure is the same as that employed in the previous study
\cite{Aoki:2006ib}. 
We use four different quark masses, two different strange quark masses
and the two lowest non-zero spacial momenta, and therefore
the total number of data points is eight for $\pi$ and $\eta$ or sixteen 
for the kaon final states.
The results obtained with the global fit using all the data are shown
in the second column in Table  \ref{tab:W0_sim_fit}.
It turns out that the simple linear function as described 
in Eq.(\ref{eq:fit1_sim}) and (\ref{eq:fit2_sim})
is in good agreement with the lattice data for all channels,
which is indicated by the reasonable $\chi^2$/dof ($\le 1.4$).
The fit results  $F_{W_0}^{\pi,\eta}(\tilde{m}^{\rm phys}_{ud},q^2)$,
$F_{W_0}^K(\tilde{m}^{\rm phys}_{ud},\tilde{m}^{\rm phys}_s,q^2)$ 
as a function of $q^2$ at the physical masses
are shown in Figs.~\ref{fig:W0_q2_R} and
\ref{fig:W0_q2_L}.

\subsection{Extrapolation to physical kinematics with sequential fitting}

\begin{figure}[tb]
\begin{center}
  \vskip -30mm
  \includegraphics[width=170mm]{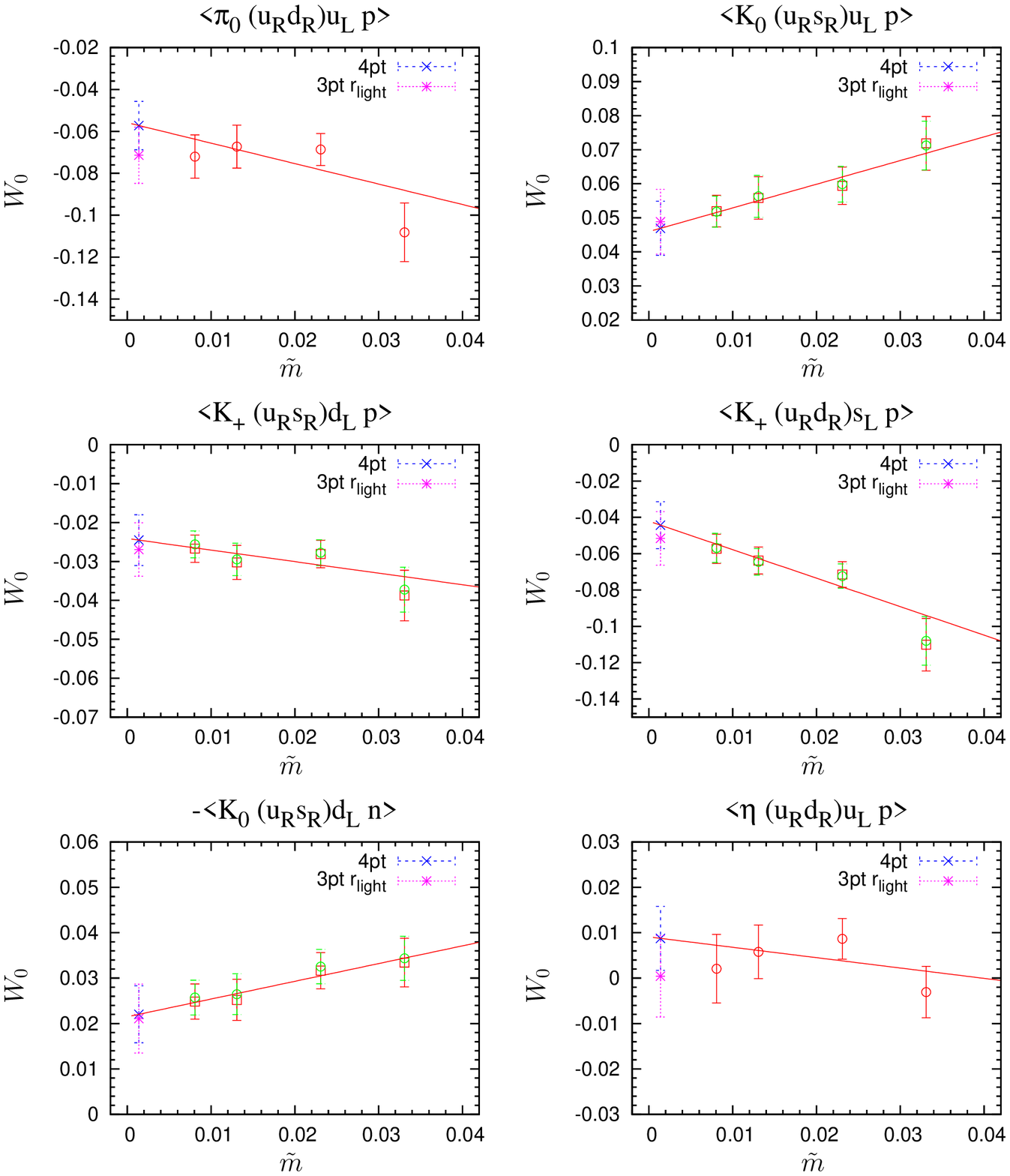}
  \vskip -20mm
  \caption{
  Results of $W_0^R(0)$ at different $\tilde m = m_{ud} + m_{\rm res}$. 
  The different open symbols shown in the matrix element of Kaon final state
  are the results at different partially quenched strange quark mass 
  $m_s=0.0343$ (circle), $m_s=0.04$ (square). 
  Straight lines show linearly fit function with all four quark masses.
  For the matrix element of $p\rightarrow K$, these are the results 
  after taking the physical strange quark mass.
   The cross symbol is the result at physical light and strange mass with 
   four fitting points and star symbol is with three fitting points
   using the range of $r_{\rm light}$ defined in the text. 
   We discuss the systematic uncertainties by using the 
   discrepancy between different fitting points 
   (for example four fitting points and three fitting points)
   in Section \ref{sec:sys_error}. 
 }
  \label{fig:W0_mpi_R}
\end{center}
\end{figure}

\begin{figure}[tb]
\begin{center}
  \vskip -30mm
  \includegraphics[width=170mm]{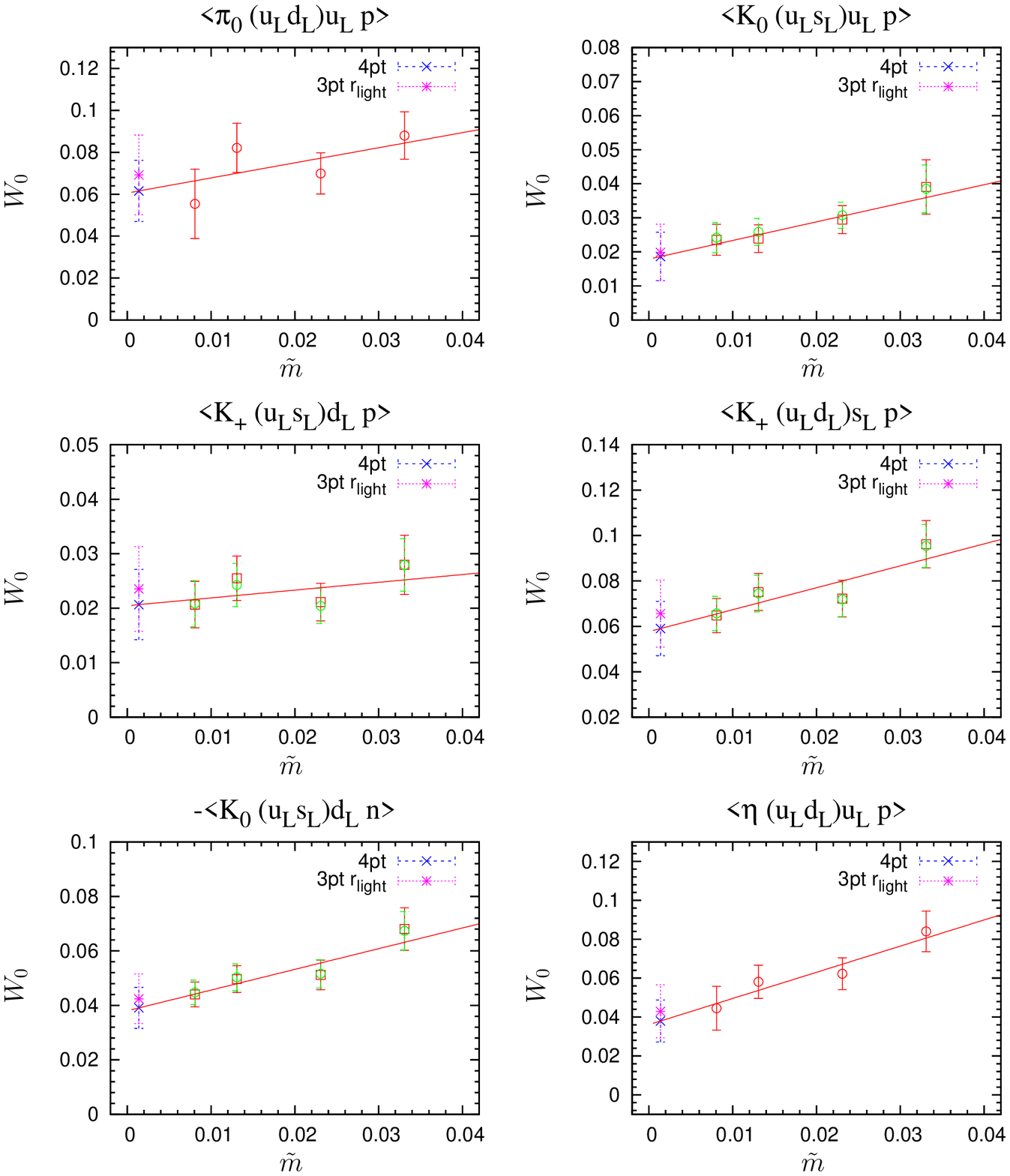}
  \vskip -20mm
  \caption{
  Results of $W_0^L(0)$ with same symbols as Fig.\ref{fig:W0_mpi_R}.
 }
  \label{fig:W0_mpi_L}
\end{center}
\end{figure}

In this procedure we first take the linear extrapolation or interpolation 
to $q^2=0$ with two spatial momentum points in each mass $\tilde{m}$
and then take a chiral extrapolation to
physical quark mass.
Figure \ref{fig:W0_mpi_R} and \ref{fig:W0_mpi_L} plot the results at $q^2=0$ point
as a function of $\tilde{m}_{ud}$ after taking the $q^2=0$ limit.
In the chiral extrapolation of the fitted data at $q^2=0$ we adopt the linear function as
\begin{eqnarray}
  f_{W_0}^{\pi,\eta}(\tilde{m}_{ud}) & = & a_0 + a_1 \tilde{m}_{ud},
  \label{eq:fit1}\\
  f_{W_0}^{K}(\tilde{m}_{ud},\tilde{m}_{s}) & = &b_0 + b_1 \tilde{m}_{ud} + b_2
   \tilde{m}_{s}, \label{eq:fit2}  
\end{eqnarray}
for the pion, $\eta$ final state or kaon final state respectively.
Here $a_i$ and $b_i$ are the free fitting parameters.
From Figure \ref{fig:W0_mpi_R} and \ref{fig:W0_mpi_L}
we observe that the linear function describes the lattice results 
quite well for each matrix elements with four different mass points,
except that the data for pion and eta in Fig.~\ref{fig:W0_mpi_R}
seems to be less consistent with the linear ansatz. 
% Actually 
%However, the $\chi^2$/dof for all matrix elements result in
%$\chi^2{\rm /dof}\lesssim 2$.
The difference of the four point fit and the three point fit
will be used in the estimate of the systematic error discussed later.
The results are shown in Table \ref{tab:W0_sim_fit} 
(see the column marked as ``Sequential'').

\subsection{Systematic errors}
\label{sec:sys_error}

The systematic errors due to using the extrapolation (or interpolation)
into physical kinematics ($q^2=0$ limit),
contribution of finite volume and
non-zero lattice spacing will be discussed in this section.
This work uses the lattice scale estimated in Ref.~\cite{Aoki:2010dy}
and the renormalization constant shown in Eq.(\ref{eq:uLL}) and Eq.(\ref{eq:uRL}).
To estimate the total error apart from the statistical error,
the systematic errors in the extrapolation, 
finite size effect and lattice artifact 
together with the error of lattice scale
and of the non-perturbative renormalization procedure,
are all added in quadrature.

At the target mass and momentum point
$(\tilde{m}_{ud},\tilde{m}_{s},q^2)
=(\tilde{m}_{ud}^{\rm phys},\tilde{m}_{s}^{\rm phys},0)$,
no chiral singularity is expected. 
Therefore, if the simulations are made closer to the target, the linear
approximation to the fitting function becomes arbitrarily precise.
However, as the simulated points might not be close enough to assume the
linearity, we need to assess the systematic error due to the choice of
this approximation.
This systematic error
is regarded as the effect of higher order than $O(\tilde{m}_{ud})$ and $O(q^2)$.
Note that the higher order effect beyond $O(\tilde{m}_{s})$ is safely
neglected as its variation around the physical point is very small
as can be estimated 
by comparing the results with $m_s=0.0343$ and $0.04$ in 
Figs.~\ref{fig:W0_q2_R} and \ref{fig:W0_q2_L}.

The main results of the relevant form factors are employed as those 
by the global fit with $0.005 \le m_{ud} \le 0.03$
(see in the second column of Table \ref{tab:W0_sim_fit}).
Note that $r$ denotes the different fitting ranges
\begin{equation}
  r_{\rm full} : [0.005,0.03],\quad r_{\rm heavy} : [0.01,0.03],\quad 
  r_{\rm light} : [0.005,0.02]
\end{equation}
which are also used in Table \ref{tab:W0_sim_fit}.
The variations of results removing the largest $\tilde{m}_{ud}$
from the global fit, 
removing the smallest $\tilde{m}_{ud}$ from the global fit and 
the result in sequential fit from the main result
provide the systematic errors coming from 
uncertainty of the fitting function for the extrapolation 
to the physical kinematics
and 
finite size effect (FSE).

The uncertainty in the extrapolation due to higher order effect 
than linearity in quark mass (and also $q^2$) is 
estimated by variance between results in $r_{\rm full}$ and $r_{\rm light}$
and variance between results with global fit and sequential fit. 
By comparing the region with and without heavy mass $m=0.03$
which is close to physical strange quark mass, we estimate the $O(\tilde m^2)$ effect. 
Furthermore since sequential fitting procedure, explained in the previous
subsection, takes into account the mass-dependence of $q^2$ slope, 
we estimate the systematic error of the extrapolation to the physical kinematics
as a part of the higher order effect, {\it e.g.} $\mathcal O(\tilde m q^2)$ terms,
beyond the $\tilde m$ and $q^2$ linear approximation  
by comparing with results in the global fit.

On the other hand the difference between results in $r_{\rm full}$ and $r_{\rm heavy}$
is expected to probe at least a part of FSE since the lightest point 
is affected most from the FSE rather than $O(\tilde m^2)$ effect.
Such estimate of FSE has been known in the calculation 
of the nucleon axial charge $g_A$
\cite{Yamazaki:2008py,Yamazaki:2009zq}
in which significant FSE was observed in the lightest quark mass
in the same gauge ensemble.
(This is also suggestive from the fact that 
the relevant form factor $W_0$ for a pion final state
is proportional to $(1+g_A)$
in the leading order of baryon chiral perturbation theory, see Ref.~\cite{Aoki:2006ib}). 
Therefore neglecting data at the lightest mass $m=0.005$ from the fitting region 
might include less contamination of FSE
(see also Fig.~10 of Ref.~\cite{Yamazaki:2009zq}). 

The systematic error including both higher order effect
$(O(\tilde m^2),\,O((q^2)^2),\,O(\tilde mq^2))$ and FSE
is evaluated by adding in quadrature the difference between the
global and sequential fitting results
in the range of $r_{\rm full}$
and the maximum difference between global fitting results 
in the range of ($r_{\rm full}$, $r_{\rm light}$)
and ($r_{\rm full}$, $r_{\rm heavy}$), even though
this procedure may be too conservative.
The magnitude is shown in the 
column denoted as ``Extrapolation'' in Table \ref{tab:W_0}.

The discretization error of $O(a)$ may arise from the inexact chiral
symmetry due to finite $L_s$.
However, as the size of the chiral symmetry breaking is small
after the additive mass shift (Eq.~\ref{eq:addm}) is performed:
$m_{\rm res}a\simeq 3\times 10^{-3}$, 
this effect can be safely neglected.
% whereas the order of $m_{\rm res}$
% $m_{\rm res}\simeq 3\times 10^{-3}$ is small. 
% Thus the chiral symmetry violation effect of DWF 
% is safely negligible.
Here the dominant discretization error at $O(a^2)$ has been estimated
using the scaling study of hadronic observable performed 
with this and finer lattice ensembles \cite{Aoki:2010dy}. 
The observed discrepancy in the spectroscopy of light meson 
(Fig.~26 in Ref.~\cite{Aoki:2010dy}) 
with the two lattice spacings is up to
1--2 \%, which amounts to about 5\% discretization error 
of the form factor $W_0$ assuming the $O(a^2)$ scaling.
We take this 5\% as the $O(a^2)$ error, 
which is more conservative than a naive 
% whose magnitude appears as the naive estimate using the 
power counting $(a\Lambda_{\rm QCD})^2\sim 0.02$ 
with $\Lambda_{\rm QCD} = 250$ MeV.

We also take into account the error 
coming from uncertainty of lattice spacing 
which is given in error of $a^{-1}=1.73(3)$ GeV
and the error of the renormalization constant which is given in
Eq.(\ref{eq:uLL}) or (\ref{eq:uRL}).

We ignore the partially quenched effect of strange quark,
which is due to the small mismatch of the sea and valence
strange masses, 
for the matrix element of $K^{+}\,,K^0$ meson final state. 
Since the valence strange quark mass dependence of $W_0$ is negligibly
small as shown in Fig.\ref{fig:W0_mpi_R} and Fig.\ref{fig:W0_mpi_L},
this effect is also negligible.
Note that we also do not consider the effect of disconnected diagrams 
in the matrix elements of the $\eta$ in the final state, but note that 
the result is valid assuming flavor SU(3) degenerate valence quark
$m^{\rm val}_{ud}=m^{\rm val}_{s}$ and ignoring partially quenched effect 
of the strange quark.

\begin{table}
\begin{center}
\caption{Table of results for renormalized $W_0^{R/L}(\mu=2{\rm GeV})$
in GeV$^2$ after global and sequential fitting. 
The error is only statistical one. 
For global fitting, we show the results with three different fitting mass-ranges,
which are all in the range $0.005\le m_{ud}\le0.03$ ($r_{\rm full}$), 
excluding the heaviest mass, $m_{ud}=0.03$, ($r_{\rm light}$) and 
excluding the lightest mass, $m_{ud}=0.005$, ($r_{\rm heavy}$).
For the sequential fitting, we show the results including all the masses.}
\label{tab:W0_sim_fit}
\begin{tabular}{c|rcrr|rc}
\hline\hline
 & \multicolumn{4}{c}{Global} & \multicolumn{2}{|c}{Sequential} \\
 matrix element & 
 \multicolumn{1}{c}{$r_{\rm full}$} & $\chi^2/$dof &
 \multicolumn{1}{c}{$r_{\rm light}$} &
 \multicolumn{1}{c}{$r_{\rm heavy}$} & 
 \multicolumn{1}{|c}{$r_{\rm full}$} & $\chi^2/$dof\\
\hline
$\langle \pi^0|(ud)_Ru_L|p\rangle$ & $-$0.103(23) & 1.4 & $-$0.132(29) & $-$0.072(34) &  $-$0.114(22) & 2.2\\
$\langle \pi^0|(ud)_Lu_L|p\rangle$ & 0.133(29) & 1.4 & 0.156(41) & 0.142(38) &  0.123(28) & 1.1\\
$\langle K^0|(us)_Ru_L|p\rangle$ & 0.098(15) & 0.4 & 0.103(19) & 0.092(29) &  0.093(15) & 0.1\\
$\langle K^0|(us)_Lu_L|p\rangle$ & 0.042(13) & 0.4 & 0.044(16) & 0.037(20) &  0.037(14) & 0.1\\
$\langle K^+|(us)_Rd_L|p\rangle$ & $-$0.054(11) & 0.8 & $-$0.060(13) & $-$0.052(21) &  $-$0.049(13) & 0.6\\
$\langle K^+|(us)_Ld_L|p\rangle$ & 0.036(12) & 0.8 & 0.040(15) & 0.041(18) &  0.041(12) & 0.6\\
$\langle K^+|(ud)_Rs_L|p\rangle$ & $-$0.093(24) & 0.6 & $-$0.108(28) & $-$0.082(39) &  $-$0.088(25) & 0.9\\
$\langle K^+|(ud)_Ls_L|p\rangle$ & 0.111(22) & 0.6 & 0.121(28) & 0.115(37) &  0.117(23) & 0.7\\
$\langle K^+|(ds)_Ru_L|p\rangle$ & $-$0.044(12) & 0.1 & $-$0.043(14) & $-$0.041(20) &  $-$0.044(12) & 0.1\\
$\langle K^+|(ds)_Lu_L|p\rangle$ & $-$0.076(14) & 0.3 & $-$0.082(17) & $-$0.076(24) &  $-$0.078(14) & 0.5\\
$\langle \eta|(ud)_Ru_L|p\rangle$ & 0.015(14) & 1.3 & $-$0.002(19) & 0.031(19) &  0.017(14) & 1.2\\
$\langle \eta|(ud)_Lu_L|p\rangle$ & 0.088(21) & 0.7 & 0.094(29) & 0.094(28) &  0.076(21) & 0.4\\

\hline\hline
\end{tabular}
\end{center}
\end{table}

\subsection{Results of proton decay matrix elements}

Table \ref{tab:W_0} summarizes the results of the relevant form factor
$W_0(q^2)$ of proton decay for all the principal matrix elements 
Eqs.~(\ref{eq:n_p_1}), (\ref{eq:n_K0})-(\ref{eq:n_eta}) at $q^2=0$.
The central values are those obtained with the global fit on $q^2$ and
the simulated quark masses for the physical kinematics
$\tilde{m}_{ud}\to\tilde{m}_{ud}^{\rm phys}$,
$\tilde{m}_s\to\tilde{m}_s^{\rm phys}$ and $q^2\to 0$, 
with the $r_{\rm full}$ range for $m_{ud}$.
The values in the first parentheses are the statistical errors.
The budget of systematic error is shown in the last four columns.
These four errors are added in quadrature to give the total systematic
error shown in the second parenthesis for each value of the form factor.

\begin{figure}[tb]
\begin{center}
  \includegraphics[width=140mm]{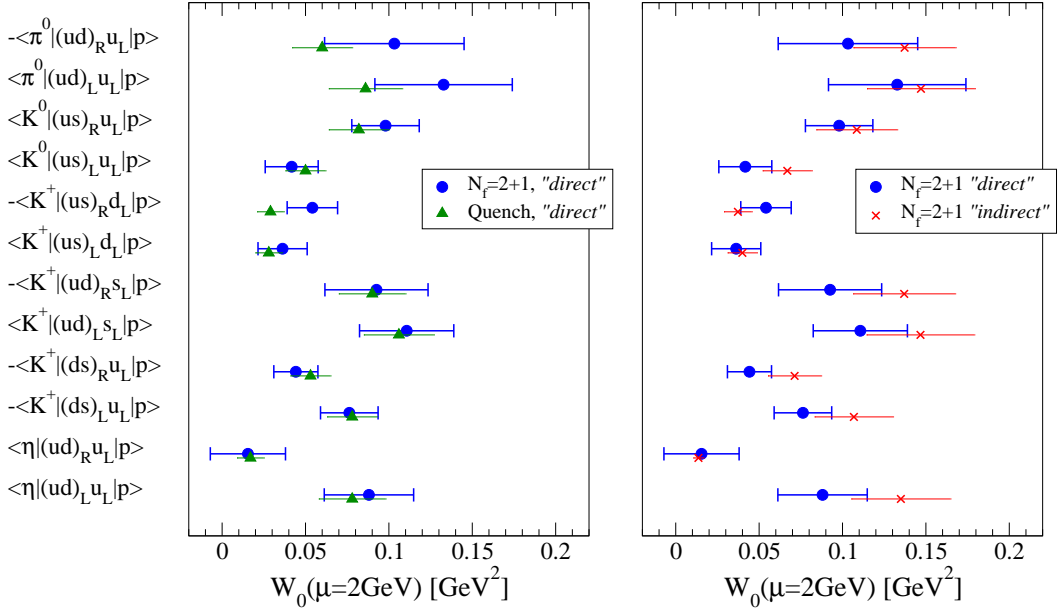}
  \vskip 3mm
  \caption{
  Summary of $W_0^{L/R}(\mu=2{\rm GeV})$ for twelve principal matrix elements. 
  Filled circles show the present results, and for the comparison the results 
  in quenched QCD (open circle) and {\it indirect} method using chiral perturbation theory 
  (cross) are plotted in the same raw. 
  }
  \label{fig:W0_sum}
\end{center}
\end{figure}

Figure \ref{fig:W0_sum} shows the results of the form factors
with the error bars expressing the total error when statistical and 
systematic errors are added in quadrature, which are marked as ``$N_f=2+1$''.
The two panels compare the results with old ones using 
some approximation. The left panel compares against the results
with quenched approximation in the {\it direct} method \cite{Aoki:2006ib}. 
The right panel shows those with the {\it indirect} method 
in the same ensembles \cite{Aoki:2008ku}. 
The sizable error for ``$N_f=2+1$'' in the current analysis 
prevents us from seeing any significant difference from the quenched 
or {\it indirect} results.
For phenomenological applications, however,
one should clearly use our $N_f=2+1$ results with the {\it direct} method 
with their total error
instead of the previous results \cite{Aoki:2006ib,Aoki:2008ku},
because each approximation previously has the systematic uncertainties which 
were not even estimated.

\begin{table}
\caption{Final results of renormalized $W_0^{L/R}(\mu=2{\rm GeV})$ 
for individual matrix elements and 
error budget of statistical and systematic uncertainties.
The first and second errors in $W_0^{L/R}$ represent statistical and systematic 
ones respectively. 
The third column denotes total error which is estimated by adding in 
quadrature statistical and systematical errors. 
The fourth column denoted as $\chi$ shows the systematic error 
of mass and momentum extrapolation/interpolation
estimated by the 
variance of extrapolation to physical kinematics and 
fifth column is uncertainties from lattice artifacts
explained in the text. 
The last two columns show the uncertainties of 
renormalization factor ($\Delta Z$) and lattice spacing ($\Delta a^{-1}$).
We also show the $p\rightarrow\pi^+\bar\nu$ decay matrix element 
using SU(2) isospin relation in Eq.(\ref{eq:piplus}).
}
\label{tab:W_0}
\begin{tabular}{rrlcccccc}
\hline\hline
 & & & Total error & \multicolumn{4}{c}{Systematic error budget}\\
 Matrix element & \multicolumn{2}{c}{$W_0(\mu=2{\rm GeV})$ GeV$^2$} 
 & (\%) & $\chi$ &
 $\mathcal O(a^2)$   & $\Delta Z$ & $\Delta a^{-1}$\\
\hline
$\langle \pi^0|(ud)_Ru_L|p\rangle$ & $-$0.103&(23) (34)& 40 & 0.033& 0.005& 0.008& 0.004\\
$\langle \pi^0|(ud)_Lu_L|p\rangle$ & 0.133&(29) (28)& 30 & 0.026& 0.007& 0.011& 0.005\\
$\langle \pi^+|(ud)_Rd_L|p\rangle$ & $-$0.146&(33) (48)& 40 & 0.047& 0.007& 0.011& 0.006\\
$\langle \pi^+|(ud)_Ld_L|p\rangle$ & 0.188&(41) (40)& 30 & 0.037& 0.010& 0.016& 0.007\\
$\langle K^0|(us)_Ru_L|p\rangle$ & 0.098&(15) (12)& 20 & 0.007& 0.005& 0.008& 0.003\\
$\langle K^0|(us)_Lu_L|p\rangle$ & 0.042&(13) (8)& 36 & 0.007& 0.002& 0.003& 0.001\\
$\langle K^+|(us)_Rd_L|p\rangle$ & $-$0.054&(11) (9)& 26 & 0.008& 0.003& 0.004& 0.002\\
$\langle K^+|(us)_Ld_L|p\rangle$ & 0.036&(12) (7)& 39 & 0.007& 0.002& 0.003& 0.001\\
$\langle K^+|(ud)_Rs_L|p\rangle$ & $-$0.093&(24) (18)& 32 & 0.016& 0.005& 0.008& 0.003\\
$\langle K^+|(ud)_Ls_L|p\rangle$ & 0.111&(22) (16)& 25 & 0.012& 0.006& 0.009& 0.004\\
$\langle K^+|(ds)_Ru_L|p\rangle$ & $-$0.044&(12) (5)& 30 & 0.003& 0.002& 0.004& 0.002\\
$\langle K^+|(ds)_Lu_L|p\rangle$ & $-$0.076&(14) (9)& 22 & 0.006& 0.004& 0.006& 0.003\\
$\langle \eta|(ud)_Ru_L|p\rangle$ & 0.015&(14) (17)& 147 & 0.017& 0.001& 0.001& 0.001\\
$\langle \eta|(ud)_Lu_L|p\rangle$ & 0.088&(21) (16)& 30 & 0.014& 0.004& 0.007& 0.003\\

\hline\hline
\end{tabular}
\end{table}

\section{Summary and Outlook}\label{sec:summary}

We have presented the lattice calculation of proton decay matrix elements 
using $2+1$ flavor dynamical domain-wall fermions, which are
essential ingredients to estimate the nucleon lifetime in grand
unified theories.
The {\it direct} method using three-point function
(nucleon)-(operator)-(meson), with non-perturbative renormalization,
was applied on a volume $L^3_\sigma\simeq 3$ fm$^3$.
Previous calculations had undermined estimate of 
systematic uncertainties on the
matrix elements at the physical kinematics.
This work made it possible to control these uncertainties for the
first time, by removing most of them, while remaining
uncertainties were given with their estimates.
The uncertainties that have been eliminated here are those 
due to the quenched approximation
\cite{Aoki:2006ib} and the use \cite{Aoki:2008ku} of the {\it indirect} method
with the tree-level baryon chiral perturbation theory.
The estimated uncertainties are the 
error in the extrapolation in quark mass and meson momentum, finite volume effect,
% combined errors from chiral extrapolation and finite volume, 
discretization error, error in the
non-perturbative renormalization and the uncertainty of the lattice scale.
The relevant form factors $W_0(q^2=0)$ of the twelve principal matrix elements
Eqs.~(\ref{eq:n_p_1}), (\ref{eq:n_K0})-(\ref{eq:n_eta}), from which 
one can calculate those for all the nucleon to pseudoscalar meson
process, has been evaluated and summarized in Table \ref{tab:W_0}
with their error estimates.
%\textcolor{red}{
% the baryon number violation process of decaying from nucleon to pseudoscalar meson, 
% has been evaluated and summarized in Table \ref{tab:W_0}
% with their error estimates.
%}

Although we have established an estimate of the proton decay 
matrix element with all the errors, 
the total errors are fairly large 
(30\%--40\% for $\pi$ final state and 20\%--40\% for the $K$ final state). 
One of the major uncertainty is the statistical error, 
especially for $p\rightarrow e^+\pi^0$ decay mode,
and that could have influenced the size of the error of combined
chiral extrapolation and finite volume effect.
A significant improvement of the current results is expected by adopting
the newly developed technique for reduction of the statistical error
\cite{Blum:2012uh}, which will be addressed in future work.
We want to emphasize, though, for now in any serious phenomenological 
application one should use the results in this study with the stated
total errors. 

\begin{acknowledgments}
We thank the members of RBC/UKQCD collaborations for their valuable
help with comments and encouragement. 
We especially would like to thank 
Peter Boyle, Paul Cooney, Chris Dawson, Luigi Del Debbio,
Taku Izubuchi, Chulwoo Jung, Adam Lichtle, 
Chris Maynard, Robert Tweedie.
Numerical calculations were performed on QCDOC computers of USQCD
collaboration and RIKEN BNL Research Center for which we thank
US DOE and RIKEN BNL Research Center.
% T.~I. and A.~S. are supported by
A.~S. is supported by
U.S. DOE contract DE-AC02-98CH10886.
This work is also supported, in part, by JSPS Kakenhi Grant
Nos.~21540289, 22224003 (YA), 
% 22540301 (TI)
and by MEXT Kakenhi Grant Nos. 23105714 (ES) %and 23105715 (TI).
.
\end{acknowledgments}

\bibliography{ndecay}

\end{document}